\documentclass[epj]{svjour}

\usepackage{graphicx}
\usepackage{axodraw}
\usepackage{fancyhdr}
\usepackage{amsmath}
\usepackage{amssymb}

\def\mytitle{My title} 
\def\myauthors{My name}  
\def\mytype{My type of session}
\def\mysession{My session}

\def\mytitle{Lepton flavor violation as a probe of squark mixing
  in supersymmetric SU(5)} 
\def\myauthors{Jae-hyeon Park}    
\def\mytype{Contributed Talk}    
\def\mysession{Flavor Physics}

\newcommand{\wt}{\widetilde}
\newcommand{\wh}{\widehat}

\newcommand{\order}{\mathcal{O}}

\newcommand{\GeV}{\ \mathrm{GeV}}

\newcommand{\MPl}{M_\mathrm{Pl}}
\newcommand{\Mgrav}{M_*}
\newcommand{\MGUT}{M_\mathrm{GUT}}

\newcommand{\tb}{\tan\!\beta}
\newcommand{\cb}{\cos\!\beta}
\newcommand{\tmg}{\tau \rightarrow \mu \gamma}
\newcommand{\teg}{\tau \rightarrow e \gamma}
\newcommand{\meg}{\mu \rightarrow e \gamma}

\newcommand{\bdg}{B \rightarrow X_d \gamma}

\newcommand{\msd}{\wt{m}_{\wt{d}}}

\newcommand{\ten}{T}
\newcommand{\fbar}{\overline{F}}
\newcommand{\adj}{\Sigma}
\newcommand{\fhiggs}{H}
\newcommand{\fbhiggs}{\overline{H}}

\newcommand{\lambdau}{\lambda_U}
\newcommand{\lambdad}{\lambda_D}

\newcommand{\deltad}{\delta^d}
\newcommand{\deltal}{\delta^l}
\newcommand{\ded}[2]{(\deltad_{#1})_{#2}}
\newcommand{\del}[2]{(\deltal_{#1})_{#2}}

\begin{document}
\title{Lepton flavor violation as a probe of squark mixing
  in supersymmetric SU(5)}
\author{Jae-hyeon Park
\thanks{\emph{Email:} jhpark@tuhep.phys.tohoku.ac.jp}%
}                     
%
%
\institute{Department of Physics, Tohoku University, Sendai 980--8578, Japan
}
%
\date{}
\abstract{
  We study flavor violation in a supersymmetric SU(5) grand unification
  scenario in a model-independent way employing mass insertions.
  We examine how the quark and the lepton sector observables restrict
  sfermion mixings.
  In many cases, a lepton flavor violating process provides
  a stringent constraint
  on the flavor structure of right-handed down-type squarks.
  In particular, $\meg$ turns out to be highly susceptible to
  the 1--3 and 2--3 mixings thereof, due to
  the radiative correction from the top Yukawa coupling
  to the scalar mass terms of $\mathbf{10}$.
  We also discuss projected sensitivity of forthcoming experiments,
  and ways to maintain the power of leptonic restrictions
  even after incorporating a
  solution to fix the incorrect quark--lepton mass relations.
} 
\maketitle
%

\section{Introduction}

A model of supersymmetry breaking/mediation, possibly in conjunction with
a model of flavor, should be compatible with the data from
flavor changing neutral current (FCNC) and $CP$ violating processes.
In particular, the past two years have seen new measurements of
$B_s$--$\overline{B_s}$ mixing \cite{DMs,phis}, which provide
new important restrictions on
the mixing between the second and the third families of down-type squarks.
On the other hand, a new experiment is going to explore the
lepton flavor violation (LFV) decay mode $\meg$,
squeezing its branching ratio down to the level of $10^{-13}$
\cite{MEG},
two orders of magnitude lower than the current upper bound.
Therefore, it can be regarded as timely to update an analysis
on supersymmetric flavor violation.

An interesting option in this style of model-independent analysis is
to work with a grand unified theory (GUT\@).
We take the SU(5) group for example.
Since a single irreducible representation contains both quarks and leptons,
their flavor structures are related.
This enables us to use both quark sector and lepton sector processes
to look into a single source of flavor violation.
It is entertaining to see which observable is supplying a tighter constraint.
The outcome can serve as a hint concerning
which sector has a higher prospect
for discovery of FCNC mediated by sparticles.
For the scalar masses and trilinear couplings to obey the GUT symmetry,
the scale of supersymmetry breaking mediation should be higher than
the GUT scale.
We suppose that this scale $\Mgrav$ is given by
the reduced Planck scale
$\MPl / \sqrt{8\pi} \sim 2 \times 10^{18}\GeV$,
or very close to it, as is the case in a gravity mediation scenario.

\begin{figure}
  \centering
\def\axoscale{0.5 }
\setlength{\unitlength}{.5pt}
\begin{picture}(290,200)(-60,20)
  \ArrowLine(-10,100)(20,100)
  \ArrowLine(20,100)(180,100)
  \DashCArc(100,100)(80,0,180){3}
  \ArrowLine(180,100)(210,100)
  \Photon(110,70)(140,30){3}{5}
  \Text(100,180)[c]{$\times$}
  \Text(43,157)[c]{$\times$}
  \Text(157,157)[c]{$\times$}
  \Text(66,170)[tl]{$\tilde{\tau}_L$}
  \Text(134,170)[tr]{$\tilde{\tau}_R$}
  \Text(30,138)[tl]{$\tilde{\mu}_L$}
  \Text(172,138)[rt]{$\tilde{e}_R$}
  \Text(-15,100)[r]{$\mu_L$}
  \Text(110,105)[b]{$\tilde{\chi}^0$}
  \Text(215,100)[l]{$e_R$}
  \Text(100,187)[b]{$(\delta^l_{33})_{RL}$}
  \Text(15,162)[b]{$(\delta^l_{23})_{LL}^*$}
  \Text(184,162)[b]{$(\delta^l_{13})_{RR}$}
\end{picture}
  \caption{Neutralino loop contribution to $\meg$ with triple mass insertions.}
  \label{fig:megtriple}
\end{figure}
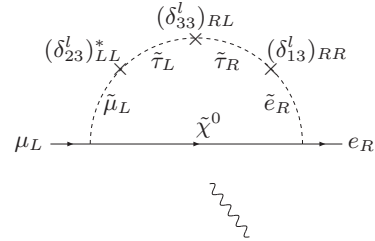

This work is by no means the first attempt in this direction.
Most notably, there is a recent article that has performed
a complete analysis in a similar framework \cite{Ciuchini:2007ha}.
Two differences are worth mentioning.
First,
we use the aforementioned $\meg$ decay mode to constrain the 1--3 and
the 2--3 mixings, in addition to $\teg$ and $\tmg$.
This seemingly unrelated process becomes relevant,
and eventually highly restrictive,
thanks to the radiative correction to the $\mathbf{10}$ representation
scalar mass matrix from the top Yukawa coupling and
the Cabibbo--Kobayashi--Maskawa (CKM) mixing.
The diagram is shown in Fig.~\ref{fig:megtriple}.
As a matter of fact, this mechanism has long been known and included
in many of the preceding model analyses.
Yet, this is the first instance of taking it into account
in a model independent analysis allowing for general flavor mixing
of sfermions, as far as we know.
Second, they assume that the quark and the lepton mass eigenstates
at the GUT scale are aligned to a high degree.
This may or may not be the case if
a solution is incorporated for fixing the wrong quark--lepton mass relations.
Especially, the first and the second families are subject to
unlimited misalignment in general \cite{Baek:2001kh}.
We propose a method to overcome this obstacle to some extent.

\section{SU(5) GUT and FCNC}

We show Yukawa couplings in the superpotential that are relevant to
the subsequent discussions.
\begin{equation}
  \label{eq:wgut}
  W \supset
  - \frac{1}{4}\epsilon_{abcde} \lambdau^{ij}
  \ten_i^{ab} \ten_j^{cd} \fhiggs^e +
  \sqrt{2}\, \lambdad^{ij} \fbhiggs_a \ten_i^{ab} \fbar_{jb} 
  .
\end{equation}
Matter fields in $\mathbf{10}$ and $\overline{\mathbf{5}}$
representations are denoted by $\ten$ and $\fbar$, respectively,
$\mathbf{5}$ and $\overline{\mathbf{5}}$ Higgses by
$\fhiggs$ and $\fbhiggs$, respectively.
The indices $a, \ldots, e$ run over components of
the fundamental representation of SU(5), and
$i, j = 1,2,3$ indicate the family.
The above Yukawa couplings, by themselves, predict mass unification
of down-type quarks and charged leptons at the GUT scale.
The third family relation is consistent with measurements
at low energies, while the first and the second are not.
One way to explain this discrepancy is to make corrections to
relatively smaller masses
by including the following non-renormalizable term:
\begin{equation}
  \label{eq:wnr}
  W_\mathrm{NR} =
  \sqrt{2} \, 
    h_2^{ij} \fbhiggs_a \ten_i^{ab} \frac{\adj_b^c}{\MPl} \fbar_{jc}
  ,
\end{equation}
where $\adj$ is the adjoint Higgs multiplet
responsible for breaking SU(5) down to the Standard Model (SM) gauge group.
We omit the other non-renormalizable terms for the sake of brevity.
This term will contribute to
the Yukawa couplings of the effective theory below the GUT scale,
expressed in terms of the SM fields as
\begin{equation}
    W_\mathrm{SSM} =
    Q^T Y_U \overline{U} H_u +
    Q^T Y_D \overline{D} H_d +
    L^T Y_E \overline{E} H_d , 
\end{equation}
where the SM fields are components of the GUT multiplets,
\begin{equation}
  \label{eq:tenfbar}
  \ten_i \simeq
  \{ Q, \overline{U}, \overline{E} \}_i, \quad
  \fbar_i \simeq
  \{ \overline{D}, L \}_i .
\end{equation}
In order to discuss flavor violation coming from the sfermion sector,
one can choose a basis of $\ten_i$ and $\fbar_i$ fields
such that
\begin{equation}
  \label{eq:initialyukawa}
  Y_U = V_Q^T \wh{Y}_U U_Q^*, \quad Y_D = \wh{Y}_D , \quad
  Y_E = U_L^T \wh{Y}_E U_R^* , 
\end{equation}
where $V_Q$ is the CKM matrix at the GUT scale,
$U_Q$, $U_L$, and $U_R$ are general unitary matrices, and
the hat on a matrix signifies that it is a diagonal matrix
with positive elements.
In this basis where $Y_D$ is diagonal,
$Y_E$ may not be diagonalized in general due to~\eqref{eq:wnr},
and it should be decomposed into the above form
using $U_L$ and $U_R$.
We can estimate
the size of an off-diagonal element of $Y_E$
in the unit of the tau Yukawa coupling,
\begin{equation}
  \frac{Y_E - \wh{Y}_D}{[\wh{Y}_E]_{33}} = 
  \frac{- \xi h_2^T}{m_\tau / (v \cb)}
  \approx - \cb \,h_2^T ,
\end{equation}
where $\xi \approx 10^{-2}$ is the ratio of $\adj$ VEV to $\MPl$, and
$v \simeq 170\GeV$ is the SM Higgs VEV\@.
Notice the suppression by the factor $\cb$ for high $\tb$.
Assuming that each element of $h_2$ is not larger than $\order(1)$,
one can see that
\begin{equation}
  \label{eq:1323mixing}
  | {[U_L]_{3a}} | , \
  | {[U_L]_{a3}} | , \
  | [U_R]_{3a}   | , \
  | [U_R]_{a3}   | \lesssim \cb ,
\end{equation}
for $a = 1,2$.
Note that the 1--3 and the 2--3 mixings are suppressed by $\cb$.
The other entries of $U_L$ and $U_R$ can be of $\order(1)$.

The SU(5) symmetry relates the soft supersymmetry breaking terms of
squarks and sleptons in a single GUT multiplet.
The scalar mass terms are given by
\begin{equation}
  \label{eq:scalarmass}
  \begin{aligned}
  - \mathcal{L}_\mathrm{soft} \supset \ &
  \fbar^\dagger m^2_{\fbar} \,\fbar +
  \ten^\dagger m^2_\ten \,\ten +
  \\
  & \fbar^\dagger \frac{\adj}{\MPl} m^{2\prime}_{\fbar} \,\fbar +
  \ten^\dagger \frac{\adj}{\MPl} m^{2\prime}_\ten \,\ten +
  \cdots ,
  \end{aligned}
\end{equation}
in which the higher dimensional terms involving $\adj$ are
suppressed by $\order(\xi)$.
 From this expression and~\eqref{eq:initialyukawa}, one can see
that the mass insertion parameters
of down-type squarks and sleptons at the GUT scale are linked by
\begin{subequations}
\label{eq:mirelations}
\begin{align}
\label{eq:mirelationRR}
  \deltal_{LL} &= U_L\,\delta^{d*}_{RR}\, U_L^\dagger
  + \order(\xi) ,
  \\
\label{eq:mirelationLL}
  \deltal_{RR} &= U_R\,\delta^{d*}_{LL}\, U_R^\dagger
  + \order(\xi) .
\end{align}
\end{subequations}

RG running from one scale down to a lower scale
generates off-diagonal elements of a scalar mass matrix.
Running from $\Mgrav$ to $\MGUT$ is
important for us to determine the boundary condition
on the soft supersymmetry breaking terms at the GUT scale.
Using one-loop approximation,
one can obtain the following form of RG contribution to
the $LL$ squark mixing at $\MGUT$,
\begin{equation}
  \label{eq:dedLLRG}
  \deltad_{LL} \simeq
  \frac{- 6}{(4\pi)^2}
  \,[V_Q^\dagger \wh{Y}_U^2 V_Q]\,
  \frac{3 m^2_0 + |A_0|^2}{\msd^2\,(\MGUT)}
  \ln \frac{\Mgrav}{\MGUT} .
\end{equation}
This comes from the CKM mixing and
the large top quark Yukawa coupling.
There can be additional $\order(\xi)$ correction, which is not written above.
However, it does not affect the validity of the following analysis
unless it cancels out the above contribution leading to a much smaller value.
That is, the left-handed squark mixing in the above expression
can be regarded as the minimal value of $\ded{ij}{LL}$
that is expected in a supersymmetric SU(5) model
with the cutoff at $\Mgrav$.
Using~\eqref{eq:mirelationLL},
one can get the right-handed slepton mixing
from~\eqref{eq:dedLLRG}.
Again, we drop the $\order(\xi)$ term in~\eqref{eq:mirelationLL},
assuming that it does not conspire with the first term to result in
a drastic cancellation.
As~\eqref{eq:1323mixing} shows that
the 1--3 and 2--3 mixings are suppressed,
one can rephrase~\eqref{eq:mirelationLL} into
\begin{equation}
  \label{eq:dela3RR}
  \del{a3}{RR} =
  [U_R]_{ab}\, \ded{b3}{LL}^* \, [U_R]_{33}^* +
  \order(\cos^2\!\beta\, \deltad_{LL}) ,
\end{equation}
for $a, b = 1, 2$,
where $[U_R]_{ab}$, the upper-left $2 \times 2$ submatrix of $U_R$,
is approximately unitary.
Keeping only the powers of $\lambda$, one can schematically write
\begin{equation}
  \label{eq:delRRCKM}
  \begin{aligned}
  \del{13}{RR} &\sim [U_R]_{11} \lambda^3 + [U_R]_{12} \lambda^2,
  \\
  \del{23}{RR} &\sim [U_R]_{21} \lambda^3 + [U_R]_{22} \lambda^2.
  \end{aligned}
\end{equation}
One finds that $\del{a3}{RR}$ is generically
not much smaller than $\lambda^3$,
unless the mixing between the first and the second families
is fine-tuned in such a way that
the two terms cancel out in either of~\eqref{eq:delRRCKM}.
For example, the mixing angle should be tuned between
$- \lambda \pm \lambda^2$
in order to have $|\del{13}{RR}| \lesssim \lambda^4$.

Before beginning numerical analysis,
one should think of how an LFV process will restrict
a squark mass insertion when we have a term like~\eqref{eq:wnr}.
Let us concentrate on $\ded{ij}{RR}$ since
an LFV process plays a particularly important role
in constraining $\ded{ij}{RR}$ rather than $\ded{ij}{LL}$ \cite{paper}.

A tau decay amplitude is dominated by the chargino loop
which is proportional to $\del{13}{LL}$ for $\teg$
or $\del{23}{LL}$ for $\tmg$.
Neglecting the $\order(\xi)$ term in~\eqref{eq:mirelationRR},
one has
\begin{equation}
  \label{eq:dela3LL}
  \del{a3}{LL} =
  [U_L]_{ab}\, \ded{b3}{RR}^* \, [U_L]_{33}^* +
  \order(\cos^2\!\beta \, \deltad_{RR}) ,
\end{equation}
for $a, b = 1, 2$,
using~\eqref{eq:1323mixing}.
The mixing between the first and the second families, parametrized by
$[U_L]_{ab}$, is not limited to be small.
This prevents us from associating $\teg$ [$\tmg$] directly with $\ded{13}{RR}$
[$\ded{23}{RR}$] in general.
Nevertheless, the following sum can be constrained
irrespective of this mixing:
\begin{equation}
  \label{eq:deltau}
  \begin{aligned}
  &|\del{13}{LL}|^2 + |\del{23}{LL}|^2 \approx \\
  &|{\ded{13}{RR}}|^2 + |{\ded{23}{RR}}|^2  
  +
  \order[\cos^2\!\beta\, (\deltad_{RR})^2] ,
  \end{aligned}
\end{equation}
which determines $B(\tau \rightarrow e / \mu\ \gamma)$.
Note that the current experimental bounds on $B(\tmg)$
and $B(\teg)$ differ only by a factor of 2.4.
Therefore, once one combines these two,
one can always give an upper bound on
each of $\ded{23}{RR}$ and $\ded{13}{RR}$, almost independent of $U_L$.
The error caused by non-vanishing 1--3 or 2--3 mixing
is diminished below 10\% even for $\tb$ as low as 3.
This permits us to do an analysis with $U_L = \mathbf{1}$,
and then consider how an LFV bound will change if
we allow for a general $U_L$ with possibly large 1--2 mixing.
If one wants to apply this conservative constraint to
the case of Fig.~\ref{fig:23RR},
the radius of the thick $\tmg$ circle should be enlarged by a factor of 1.9.
The thick $\teg$ circle in Fig.~\ref{fig:13RR}
should be expanded by a factor of 1.2.
Similarly, the future bounds can be modified:
multiply each by $\sqrt{2}$.

Unlike the tau decay modes, $\meg$ is more involved.
The dominant contribution comes from the triple insertion graph
in Fig.~\ref{fig:megtriple}.
The decay rate is proportional to
  \begin{align}
  \label{eq:megpropto}
    & |\del{13}{RR} \del{32}{LL}|^2 + |\del{13}{LL} \del{32}{RR}|^2 
    \gtrsim \\
    & \min \{ |\del{13}{RR}|^2, |\del{23}{RR}|^2 \} \cdot
  [ |{\ded{13}{RR}}|^2 + |{\ded{23}{RR}}|^2 ] , \nonumber
  \end{align}
ignoring the term suppressed by $\cos^2\!\beta$.
We learned in~\eqref{eq:delRRCKM}
that the first factor on the right-hand side is at least
as large as $\ded{13}{LL}$
unless the mixing angle in $[U_R]_{ab}$ is fine-tuned.
Thus, the $U_L$-independent upper bound on each of
$\ded{13}{RR}$ and $\ded{23}{RR}$, should be
given by a $\meg$ ring in Fig.~\ref{fig:23RR}.
That is, Fig.~\ref{fig:23RR} is not modified even with this conservative
interpretation, 
while the $\meg$ circles in Fig.~\ref{fig:13RR}
should be replaced by those in Fig.~\ref{fig:23RR}.

\section{Constraints on scalar mixings}

We perform a numerical analysis for
the two cases displayed in Table~\ref{tab:insertions},
with $U_L$ and $U_R$ set to a unit matrix.
\begin{table}
  \renewcommand{\arraystretch}{1.1}
  \begin{tabular}{cccccc}
    \hline
    Fig. & 
    $12LL$ & $13LL$ & $23LL$ & $13RR$ & $23RR$ \\
    \ref{fig:23RR} & 0.000048 &
    0.0015 & 0.0074 & 0 & free            \\
    \ref{fig:13RR} & 0.000048 &
    0.0015 & 0.0074 & free & 0               \\
    \hline
  \end{tabular}
  \caption{Values of mass insertion parameters to be given as
    boundary conditions at the GUT scale.
    The phase of a fixed $\ded{ij}{LL}$ is equal to
    $\arg( - V_{ti}^* V_{tj} )$,
    as can be expected from~\eqref{eq:dedLLRG}.}
  \label{tab:insertions}
\end{table}%
A parameter indicated as `free' is a variable to be scanned over,
and the other three are fixed at the respective specified numbers,
expected from RG running between $\Mgrav$ and $\MGUT$.
A case with $\ded{ij}{LL}$ as a free parameter,
is included as well in~\cite{paper}.
We use the scalar mass $m_0 = 220\GeV$ and the gaugino mass
$M_{1/2} = 180\GeV$ at the GUT scale, which cause
both the down-type squark masses and the gluino mass to be 500 GeV
at the weak scale.
We choose $\tb = 5$ and positive $\mu$.
Treatment of the experimental data is summarized in~\cite{paper}.

\begin{figure}
  \centering
{\includegraphics[height=57mm]{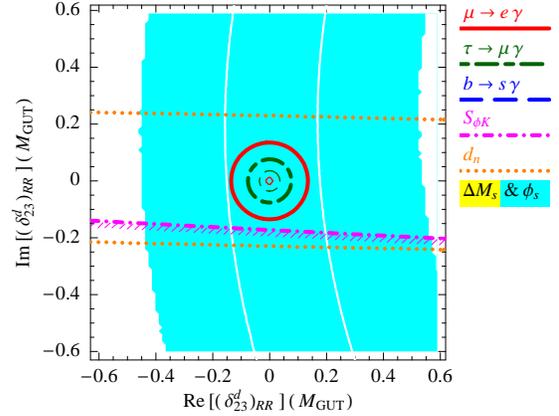}}
\caption{Constraints on the complex plane of $\ded{23}{RR}$,
    with $\ded{ij}{LL}$ generated from
    RG running between the reduced Planck scale and the GUT scale.
    For each LFV process,
    the thick circle is the present upper bound and
    the thin circle is the prospective future bound.
    A gray (cyan) region is consistent with
    $\Delta M_{B_s}$ and $\phi_{B_s}$,
    given 30\% uncertainty in the $\Delta B = 2$ matrix element.
    The white curves mark a possible improved
    constraint from $\Delta M_{B_s}$
    with 10\% hadronic uncertainty.
    Of the two sides of the $S_{\phi K}$ curve,
    the excluded one is indicated by the thin short lines.}
\label{fig:23RR}       
\end{figure}

Result of the first case is
displayed in Fig.~\ref{fig:23RR}.
It indicates regions of $\ded{23}{RR} = \del{23}{LL}^*$ (in)consistent with
observations.
The gray (cyan) belt is allowed by $\Delta M_{B_s}$ and $\phi_{B_s}$.
However, most of it is ruled out by the LFV processes.
One can guess that this should be the case even in the near future,
comparing the belt between two white curves and the thin circles
with their centers at the origin.
In particular, the $\meg$ data from the MEG experiment
should be able to kill all the parameter space except for the tiny disk
around the origin.
This is because our default value of
a mass insertion is the one which is minimally expected from
RG running.
We believe that our choice of $LL$ squark mass insertions
are more reasonable than zero
in a scenario where the soft terms are generated
around the Planck scale such as gravity mediation.
It should also be mentioned that
an RG-induced $LL$ insertion is not only critical to $\meg$, but also
to $B_s$--$\overline{B_s}$ mixing \cite{Endo:2006dm}.
Indeed, the presence of $\ded{23}{LL}$ is rendering
the $\Delta M_{B_s}$ constraint on $\ded{23}{RR}$ tighter,
though not as tight as the leptonic ones.
Further discussions on this plot and those with different parameters
can be found in~\cite{paper}.

\begin{figure}
{\includegraphics[height=57mm]{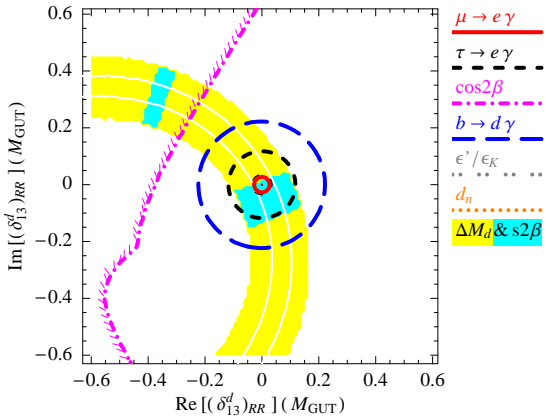}}
\caption{Constraints on the complex plane of $\ded{13}{RR}$,
  obtained in the same way as Fig.~\ref{fig:23RR}.
  Of the two sides of a $\cos 2\beta$ curve,
  the excluded one is indicated by the thin short lines.}
\label{fig:13RR}       
\end{figure}

We present in Fig.~\ref{fig:13RR},
constraints on $\ded{13}{RR} = \del{13}{LL}^*$.
The light gray (yellow) belt is compatible with $\Delta M_{B_d}$,
which is further reduced by $\sin 2\beta$ into the two
gray (cyan) regions.
One of them is excluded by the $\cos 2\beta$ measurement.
The resulting area is completely consistent with $\bdg$.
The width of this area is comparable to the diameter of
the circle from $\teg$.
The restriction from $\meg$ is so strong that it rules out
most of the gray zone.
The $\meg$ disk in this plot is smaller than that in Fig.~\ref{fig:23RR}.
The reason is that the decay amplitude is proportional to
$\del{23}{RR} \sim \lambda^2$
here, but to $\del{13}{RR} \sim \lambda^3$ there.
In a few years, improved lattice QCD should be able to
narrow the $\Delta M_{B_d}$ belt
down to the one between the two white curves,
whose width is again comparable to the diameter of the future $\teg$ disk.
The MEG constraint is so tight that it
appears to be a single dot at the origin.

\section{Conclusion}

We imposed hadronic and leptonic constraints
on sfermion mixing in a class of supersymmetric models
with SU(5) grand unification.
We did not particularly assume that
the sfermion mass matrices have a universal form at any scale,
but rather that any off-diagonal entry may be nonzero,
which is generically the case in gravity mediated supersymmetry breaking.
Those off-diagonal elements are encoded by the
dimensionless mass insertion parameters,
in terms of which we express
experimental bounds on flavor non-universality at the GUT scale.
The most restrictive constraint on a given mass insertion may change
from one process to another
if one modifies a parameter or adopts a different scheme of
uncertainty treatment.
A projected bound is also subject to variation depending on
the estimated performance of a future experiment.
Nevertheless, it should be obvious that an LFV decay mode provides
an essential limit on a sfermion mixing in many cases.
This is true even when one introduces non-renormalizable terms
to accommodate the lighter down-type quark and charged lepton masses.
In particular, the apparently unrelated mode $\meg$
turns out to be remarkably sensitive to a mixing involving the third family.
This sensitivity is will be much higher
with the progress of the MEG experiment.

\section*{Acknowledgments}

The author thanks P.~Ko and Masahiro~Yamaguchi
for the collaboration.
He was supported by the JSPS postdoctoral fellowship program
for foreign researchers and the accompanying grand-in-aid no.\ 17.05302.

%
%

\end{document}